\documentclass{jfm}
\usepackage{graphicx}
\usepackage{psfrag}
\usepackage{tikz}
\usepackage{amsmath}
\usepackage{subcaption}
\usepackage{gnuplottex}

\DeclareRobustCommand{\dashedline}{\raisebox{2pt}{\tikz{\draw[black,dash pattern=on 2pt off 1.5pt,line width=0.9pt](0,0) -- (7mm,0)}}}
\DeclareRobustCommand{\dottedline}{\raisebox{2pt}{\tikz{\draw[black,dash pattern=on 0.7pt off 0.7pt,line width=0.9pt](0,0) -- (7mm,0)}}}
\DeclareRobustCommand{\solidline}{\raisebox{2pt}{\tikz{\draw[black,solid,line width=0.9pt](0,0) -- (7mm,0)}}}

\shorttitle{Kelvin-Helmholtz billows above Richardson number $1/4$}
\shortauthor{J. P. Parker, C. P. Caulfield and R. R. Kerswell}

\title{Kelvin-Helmholtz billows above Richardson number $1/4$}

\author{J. P. Parker\aff{1}
  \corresp{\email{jpp39@cam.ac.uk}},
  C. P. Caulfield\aff{2,1}
 \and R. R. Kerswell\aff{1}}

\affiliation{\aff{1}Department of Applied Mathematics and Theoretical Physics,
Centre for Mathematical Sciences, University of Cambridge, Wilberforce Road, Cambridge CB3 0WA, UK
\aff{2}BP Institute, University of Cambridge, Madingley Road, Cambridge CB3 0EZ, UK}

\begin{document}
\maketitle

\begin{abstract}
    We study the dynamical system of a forced stratified mixing layer at finite
    Reynolds number $Re$, and Prandtl number $Pr=1$.
    We consider a
    hyperbolic tangent background velocity profile
    in the two cases of hyperbolic tangent and uniform background buoyancy stratifications.
    The system is forced in such a way that these background profiles are a steady
    solution of the governing equations.
    As is well-known, if the minimum gradient Richardson number of the flow, $Ri_m$, is less than a certain critical value $Ri_c$, the flow is
    linearly unstable to Kelvin-Helmholtz instability in both cases. Using Newton-Krylov iteration, we find
    steady, two-dimensional, finite amplitude elliptical vortex
    structures, i.e.
    `Kelvin-Helmholtz billows', existing above $Ri_c$. Bifurcation diagrams are produced using branch continuation,
    and we explore how these diagrams change with varying $Re$.
    In particular, when $Re$ is sufficiently high we find that
    finite amplitude Kelvin-Helmholtz billows exist at $Ri_m>1/4$, where the flow is linearly stable by the Miles-Howard theorem.
    For the uniform background stratification, we give a simple explanation of the dynamical system,
    showing the dynamics can be understood on a two-dimensional manifold embedded in state space,
    and demonstrate the cases in which the system is bistable.
    In the case of a hyperbolic tangent stratification, we also describe a new, slow-growing, linear instability
    of the background profiles at finite $Re$, which complicates the dynamics.
\end{abstract}

\section{Introduction}
The Miles-Howard theorem \citep{miles_stability_1961, howard_note_1961} tells us that for inviscid, infinitesimal perturbations to steady,
one-dimensional, parallel shear flows, the minimum gradient Richardson number $Ri_m$ of the flow must be less
than $1/4$ for such `linear' perturbations to grow exponentially. From this, it is often argued that oceanic measurements will always
find a Richardson number greater than or equal to $1/4$, otherwise turbulence will ensue \citep[see][and references therein]{Smyth2019},
despite the very specific restrictions on the applicability of the theorem. In this paper
we will examine two aspects of these restrictions, namely that perturbations are infinitesimal,
and that $Re=\infty$.

With finite amplitude perturbations, nonlinear effects can no longer be neglected. There is various
evidence that for flows susceptible to Kelvin-Helmholtz instability (KHI), complex nonlinear behaviour
exists when $Ri_m > 1/4$. \citet{kaminski_nonlinear_2017} showed that perturbations which grow transiently before decaying
in the linearised setting can lead to turbulent-like irreversible mixing with $Ri_m>1/4$ when nonlinearity is included.
\citet{howland_testing_2018} showed that as $Ri_m\to1/4$ from below, the maximum amplitude
of a saturated Kelvin-Helmoltz billow does not tend to zero, but to some finite value.
One possible cause of these observations is that the pitchfork bifurcation, generically expected
to occur at the critical Richardson number, $Ri_c$, is subcritical, so that finite amplitude states exist
above $Ri_c$. This could mean that the system is bistable in a certain range of $Ri_m$ with $Ri_m>Ri_c$.
(Note that by `subcritical' here we mean those regions where the laminar flow is
linearly stable, \textit{above} $Ri_c$, consistent with normal dynamical systems terminology,
as opposed to the occasional oceanographic usage meaning \textit{below} $Ri_c$.)

Historically, the best way to determine the nature of the bifurcation has been to consider the next order
nonlinear effects, a so-called weakly nonlinear analysis. Such analysis was performed by \citet{maslowe1977}
and \citet{brown1981} for stratified shear layers, finding subcriticality (in the above sense) for $Pr<1$.
However, our results suggest that the weakly nonlinear analysis can potentially be misleading, as discussed
in section \ref{sec:conclusion}, since higher order effects can quickly dominate.

 More recently, as it has become possible computationally to solve the Navier-Stokes equations
directly, finding the finite amplitude states which arise from bifurcations has emerged
as an alternative. Newton's method can be used to find solutions of nonlinear problems,
such as steady states, iteratively.
The introduction of Newton-Krylov methods \citep{edwards1994krylov},
where a Krylov-subspace method such as generalised minimal residuals (GMRES) \citep{saad1986gmres} is used to solve the
linear system inexactly at each Newton step, has allowed this to be applied to very high dimensional systems
for which it is prohibitively expensive to work with matrices directly \citep[for a comprehensive review, see][]{dijkstra_2014}.
It is also possible to use Newton's method to find and track bifurcation points of high dimensional
dynamical systems \citep{salinger2002,haines_hewitt_hazel_2011}. \citet{netsanchez2015} used a matrix-free
bifurcation tracking technique with a Newton-Krylov method, as employed in this paper,
and further extended this to find bifurcations of periodic orbits.

In this paper,
we find the exact coherent states that bifurcate from the laminar flow at $Ri_c$, and track these
as both $Ri_m$ and $Re$ vary, to build a picture of the dynamical system
near $Ri_m=1/4$, and, crucially, answer the question of whether the system can be bistable above
$Ri_c$.
Two different models susceptible to KHI are considered.
The first, the `Holmboe' model \citep{holmboe_1962}, with a hyperbolic tangent buoyancy profile,
is the standard model in this field \citep{hazel1972numerical,klaassen_evolution_1985, smyth_instability_1991, Mallier2003},
but we demonstrate that complex behaviour
arises--associated with what we believe to be a previously unreported linear instability--and
 dominates at long times when this model
is forced onto the system at finite $Re$, obscuring the KHI. We then examine an alternative `Drazin' model \citep{drazin_1958},
with a uniform stratification, which shares many of the features of the Holmboe model but does
not exhibit this complex behaviour. Note that, with the parameters studied, both models
are only known to be susceptible to stationary KHI,
and not the propagating Holmboe wave instability.
The paper proceeds as follows: in section \ref{sec:methods}, we describe the methodology and code
used. In section \ref{sec:tanh} a bifurcation diagram is presented for the Holmboe model,
as well as a description of the newly discovered linear instability. In section \ref{sec:linear}, a bifurcation diagram
and a full description of the dynamics is given for the Drazin model. Section \ref{sec:conclusion}
gives a brief discussion of these results.

\section{Methodology}
\label{sec:methods}
We consider the Boussinesq equations in two dimensions, and study the nonlinear evolution
of perturbations away from a steady parallel velocity profile
$U(z)$ and buoyancy stratification $B(z)$. Solving for the perturbation away from these constant-in-time
profiles is equivalent to solving for the full system, with an artificial body force to counteract
diffusion.
In non-dimensional form, the equations are:
\begin{align}
    \label{eq:perturbation1}
    \partial_t {u} + \left(U+{u}\right) \partial_x u + {w} \partial_z \left(U+{u}\right) &= -\partial_x {p}
    + \frac{1}{Re}\left(\partial_x^2 {u} + \partial_z^2 {u} \right), \\
    \partial_t {w} + \left(U+{u}\right) \partial_x {w} + {w} \partial_z {w} &= -\partial_z {p}
    + \frac{1}{Re}\left(\partial_x^2 {w} + \partial_z^2 {w} \right) + Ri_b {b}, \\
    \partial_t {b} + \left(U+{u}\right) \partial_x {b} + {w} \partial_z \left(B+{b}\right) &=
    \frac{1}{PrRe}\left(\partial_x^2 {b} + \partial_z^2 {b} \right), \\
    \label{eq:perturbation4}
    \partial_x {u} + \partial_z {w} &= 0.
\end{align}
Here $u$ is the fluid velocity in the horizontal ($x$) direction, and $w$ is the velocity in the vertical ($z$)
direction. Buoyancy acts in the positive $z$ direction. We impose periodic boundary conditions
at $x=0$ and $x=L_x$, and at $z=\pm L_z$ we enforce no-penetration ($w=0$), stress-free ($\partial u/\partial z=0$),
and insulating ($\partial b/\partial z=0$) boundary conditions.
Given the dimensional shear layer depth, $2L$, velocity difference $2\Delta U$, density difference $2\Delta \rho$, typical density $\rho^*$,
and diffusivities of momentum $\nu$ and density $\kappa$, the Reynolds number is defined as $Re=\frac{\Delta U L}{\nu}$,
the Prandtl number $Pr=\frac{\nu}{\kappa}$, and the bulk Richardson number $Ri_b=\frac{g}{\rho^*}\frac{L\Delta\rho}{\Delta U^2}$.
Throughout, we take $Pr=1$ for simplicity.
Two different choices of $U$ and $B$ are considered in
sections \ref{sec:tanh} and \ref{sec:linear} respectively.
For both background flows studied, the minimum gradient Richardson number $Ri_m$, as relevant to the Miles-Howard theorem,
is equal to the bulk Richardson number $Ri_b$.

\subsection{Time-stepping}
A new solver was developed to solve the Boussinesq equations around arbitrary background
flows. Time integration uses a third order Runge-Kutta-Wray scheme, and spatial derivatives are handled
pseudo-spectrally in the periodic horizontal direction, and with explicitly conservative quasi-second
order finite differences in the vertical, on a non-uniform staggered grid with more points closer to the central
shear layer. The code was validated against DIABLO \citep{taylor2008}.
Further, a linearised version of the same timestepper was produced, and validated against very low amplitude
states in the full nonlinear solver.
For the system studied in section \ref{sec:tanh}, a grid is used with 256 equispaced points in the streamwise direction,
and 512 points in the vertical direction, with a greater density of points in the middle of the domain.
For the system studied in section \ref{sec:linear}, 128 points are used in the streamwise direction,
covering a shorter domain, and 768 vertically, in order to accurately capture behaviour at higher $Re$.
The results are validated by reconverging certain solutions at a higher resolution of
$384\times768$ in section \ref{sec:tanh}
and $256\times1024$ in section \ref{sec:linear}.

\subsection{Steady states and bifurcation points}
\label{sec:tracking}
Formally, we may describe our dynamical system as the evolution of a state $X$ by a time $t$ through
\begin{equation}
    X(t_0+t)=F\left(X(t_0),t;Ri_b,Re\right),
\end{equation}
where $Ri_b$ and $Re$ are the constant parameters at which we are considering the evolution.
Finding steady states of the flow is then equivalent to finding solutions to
\begin{equation}
    \label{eq:steadystate}
    F(X,T;Ri_b,Re)-X = 0
\end{equation}
for some arbitrary fixed $T$. A larger $T$ acts to precondition the equations, but if it is too large,
computation will be prohibitively expensive. For our system, we found $T=11$ to be a good compromise.
It is possible, though extremely unlikely, that this will also
find a periodic orbit of period $T$.

Solving (\ref{eq:steadystate}) is done by using Newton-GMRES (generalised minimum residual)
iteration on an initial guess.
Our implementation closely matches that employed by \citet{chandler_kerswell_2013}, including the use
of a trust region to make the algorithm globally convergent. The GMRES iteration at each Newton step
is continued until the residual is less than $10^{-2}$, and the Newton iteration is continued until
its residual, the norm of the left-hand side of (\ref{eq:steadystate}), is less than $10^{-8}$.
Through trial and error, we converge a steady state solution, the result of a very long time integration
of equations (\ref{eq:perturbation1}-\ref{eq:perturbation4}), at $Re=1000$ and $Ri_b=0.2$, in both the
flows studied in this paper.
Once one state is found at these particular $Ri_b$ and $Re$, we converge another very close by
at a different $Ri_b$ but the same $Re$. We then follow the solution branch at this $Re$ over a range of $Ri_b$ using
pseudo-arclength continuation \citep{keller1977applications}. We examine the stability of the branch with
Arnoldi iteration, using a linearised version of the same timestepping code.

The stability analysis
reveals the existence of bifurcation points, where eigenvalues of the state cross a stability boundary.
To continue these bifurcation points to different $Re$, we use the states found by stability analysis
as an initial guess in a different iterative solver. The system we solve is similar to that implemented
in LOCA \citep{salinger2002}, but we use a matrix free method, as discussed in detail in \citet{Umbria2016}.
We look for solutions to
\begin{subequations}
    \label{eq:bifurcpoints}
\begin{align}
        F(X,T;Ri_b,Re)-X &= 0,\\
        F_X(X,Y,T;Ri_b,Re)-Y &= 0,\label{eq:enforceeigenmode}\\
        Y\cdot A -1 &= 0,\label{eq:normalise}
\end{align}
\end{subequations}
with Newton-GMRES. In this case we allow $X$, $Y$ and $Ri_b$ to be found by the iteration, but hold $Re$ fixed.
Here $F_X(X,Y,t;Ri_b,Re)$ is the linearised time evolution of a state $Y$ about a nonlinear state $X$,
computed using the linearised timestepper. Equation (\ref{eq:enforceeigenmode}) enforces that $Y$ is a neutral eigenmode
of the Jacobian at $X$.
We normalise $Y$ using (\ref{eq:normalise}), with some fixed arbitrary state $A$.
Once bifurcation points are found at a particular $Re$, they are reconverged at higher $Re$.
We are particularly interested in how the $Ri_b$ value of the bifurcation point varies with $Re$.

Equations (\ref{eq:bifurcpoints}) find bifurcation points with purely real neutral eigenmodes, i.e. pitchfork
and saddle-node bifurcations. For Hopf bifurcations, a set of five equations is needed, including
two different linearised time evolutions. These arise from the real and imaginary parts of the eigenvalue $e^{i\theta}$.
The following are solved for the unknowns $X$, $Y_1$, $Y_2$, $Ri_b$ and $\theta$:
\begin{subequations}
    \label{eq:bifurchopf}
    \begin{align}
        F(X,T;Ri_b,Re)-X &= 0,\\
        F_X(X,Y_1,T;Ri_b,Re)-\cos{\theta}\;Y_1+\sin{\theta}\;Y_2 &= 0,\\
        F_X(X,Y_2,T;Ri_b,Re)-\sin{\theta}\;Y_1-\cos{\theta}\;Y_2 &= 0,\\
        Y_1\cdot A -1 &= 0,\\
        Y_2\cdot A &=0.
    \end{align}
\end{subequations}
The additional computational requirements of (\ref{eq:bifurchopf}) mean that we are unable to track Hopf
bifurcations to as high Reynolds numbers as pitchfork and saddle-node bifurcations.

\section{Results}
\subsection{Hyperbolic tangent stratification: the Holmboe model}
\label{sec:tanh}

\begin{figure}
    \begin{subfigure}{0.49\textwidth}
        \begin{gnuplot}[terminal=epslatex, terminaloptions={size 7cm,5cm}]
            set key off
            set size 1, 1
            set xtics out nomirror 0,0.002,0.25
            set ytics out nomirror 0,0.2,1
            set border lw 0.5
            set xrange [0.243:0.2512]
            set yrange [0:0.75]
            set xlabel "$Ri_b$"
            set ylabel "$\\|X\\|$" offset 2,0 rotate by 0

            set style line 1 lw 3 lc 7

            set multiplot

            set label "$Ri_c$" at graph 0.72,-0.07
            set label "$Ri_h$" at graph 0.06,0.82
            set label "$Ri_s$" at graph 0.89,0.32

            plot \
            "-" w l lt 2 lw 1,\
            "fig1a/upperleft" w l ls 1 lt 1,\
             "fig1a/uppermid" w l ls 1 lt 6,\
             "fig1a/upperright" w l ls 1 lt 4,\
             "fig1a/midbranch" w l ls 1 lt 2,\
             "fig1a/lowerbranch" w l ls 1 lt 1,\
             "fig1a/basestate" w l ls 1 lt 2,\
             "-" pt 7 lc rgb "green",\
             "-" pt 7 lc rgb "blue",\
             "-" pt 7 lc rgb "red",\
             "-" w p pt 1 lc rgb "black"
             0.25 0
             0.25 1
             e
             0.249414 0
            e
            0.250176 0.249405
            e
            0.243921	0.672592
            e
            0.245688 0.604233
            0.249669 0.0890092
            0.248799 0.447446
            e

            unset label

            set origin 0.12,0.2
            set size 0.65,0.55
            set xrange [0.2493:0.2502]
            set yrange [0:0.3]
            unset xlabel
            unset ylabel
            set xtics out nomirror 0,0.0005,0.25 offset 0,0.5,0
            set ytics out nomirror 0,0.2,1 offset 0.5,0,0

            plot \
            "fig1a/upperleft" w l ls 1 lt 4,\
             "fig1a/uppermid" w l ls 1 lt 6,\
             "fig1a/upperright" w l ls 1 lt 4,\
             "fig1a/midbranch" w l ls 1 lt 2,\
             "fig1a/lowerbranch" w l ls 1 lt 1,\
             "fig1a/basestate" w l ls 1 lt 2,\
             "-" w p pt 1 lc rgb "black"
             0.249669 0.0890092
             e

            unset multiplot
        \end{gnuplot}
        \raisebox{4.2cm}[0pt][0pt]{
            \parbox{10cm}{\caption{}\label{fig:re1000tanh}}
        }
        \vspace{-1cm}
    \end{subfigure}
    \begin{subfigure}{0.49\textwidth}
        \psfrag{10m3}{$10^{-3}$}
        \psfrag{10m4}{$10^{-4}$}
        \psfrag{10m5}{$10^{-5}$}
        \psfrag{0}[cc][cc][1][90]{$0$}
        \psfrag{0.1}[cc][cc][1][90]{$0.1$}
        \psfrag{0.15}[cc][cc][1][90]{$0.15$}
        \psfrag{0.6}[cc][cc][1][90]{\color[RGB]{150, 150, 150}$0.6$}
        \psfrag{0.7}[cc][cc][1][90]{\color[RGB]{150, 150, 150}$0.7$}
        \includegraphics[width=\textwidth]{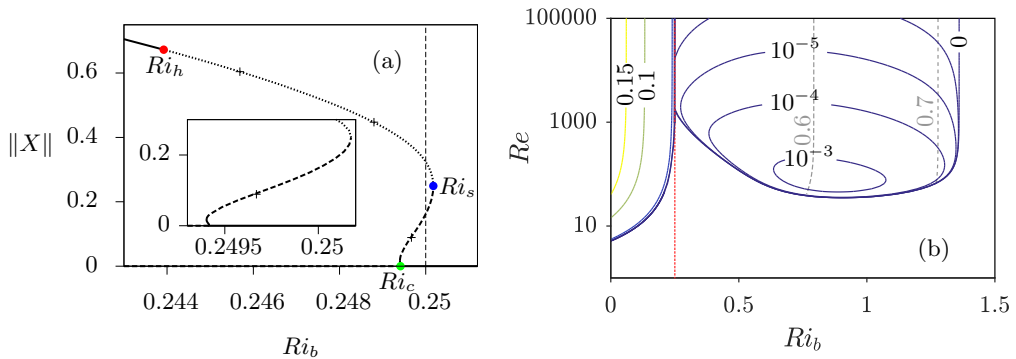}
        \raisebox{1.5cm}[0pt][0pt]{
            \parbox{11cm}{\caption{}\label{fig:tanhstab}}
        }
        \vspace{-1cm}
    \end{subfigure}
    \caption{(a) Bifurcation diagram for the flow with hyperbolic tangent background stratification, at $Re=4000$,
    showing the variation of $\|X\|$ over a (very narrow) range of $Ri_b$.
                 At different points, the solution branch has
              one stationary unstable direction (\dashedline),
              two stationary unstable directions (\dottedline),
              or is stable to stationary disturbances (\solidline). All states in this diagram
              are unstable to the new, propagating instability.
              $Ri_h$ is plotted with a red dot, $Ri_s$ with a blue dot and $Ri_c$ with a green dot.
              The crosses mark points converged at the higher resolution of $384\times768$.
              (b) Contours of growth rate from a linear stability analysis of the background flow.
              The dashed grey lines show contours of imaginary part.
              For $Ri_b<1/4$, the dominant instability mechanism is KHI, with
              a purely real growth rate. The newly described instability, discussed in the text,
              is the only one for $Ri_b>1/4$, and has a very small growth rate, with nonzero imaginary part.}
\end{figure}

First we consider a background profile of $U = \tanh{z}$, $B = \tanh{z}$. This is a commonly used
model of a mixing layer, introduced by \citet{holmboe_1962}. It has the useful property that, at infinite
$Re$, the linear stability analysis can be performed analytically by hand \citep{miles_1963}.
With this choice, we find that the minimum gradient Richardson number $Ri_m$ is equal to $Ri_b$,
and so the Miles-Howard theorem tells us that the flow is certainly stable for $Ri_b>1/4$.
We choose $L_x=4\pi$, which is one wavelength of the most unstable mode at $Ri_b=1/4$ as $Re\to\infty$, assuming a domain of infinite height vertically.
We take $L_z=10$. This is a compromise between being large enough that the boundaries do not
significantly affect the flow but small enough to keep computation costs down. The finite value of $L_z$ means that
 $Ri_c$ tends to a value slightly less
than $1/4$ as $Re\to\infty$.

Following \citet{howland_testing_2018} we define the energy of perturbations to be
\begin{equation}
    E = \frac{1}{2L_x}\int_0^{L_x} \mathrm{d}x \int_{-L_z}^{L_z} \mathrm{d}z \left({u}^2 + {w}^2 + Ri_b {b}^2\right).
\end{equation}
State space is taken as the space of all possible incompressible perturbation flows $X=\left({u},{w},{b}\right)$,
with norm $\left\|X\right\| := \sqrt{2E}$.
Note that $p$
is not a dynamical variable as it can be calculated from a Poisson equation forced by the velocity field.


Figure \ref{fig:re1000tanh} shows a bifurcation diagram at $Re=4000$. Where the background state
becomes unstable to KHI at $Ri_c\approx 0.2494$,
a pitchfork bifurcation occurs (the green dot on figure \ref{fig:re1000tanh}), giving rise to a branch of
finite amplitude, billow-like states. This branch is initially
stable--except to the unrelated instability discussed below--and decreasing in $Ri_b$,
but there is soon a saddle-node bifurcation (see inset in figure \ref{fig:re1000tanh}) and it then increases in $Ri_b$.
As the unstable branch increases in amplitude, $Ri_b$ increases, and we find steady, though unstable, states above $Ri_b=1/4$.
There is another saddle-node bifurcation at $Ri_s$ (blue dot), adding a second
unstable direction to the branch.
The Hopf bifurcation, at $Ri_h\approx 0.244$ (red dot), has two neutral eigenmodes which are connected
with the eigenmodes of the two saddle-node bifurcations via a periodic orbit.

\begin{figure}
        \begin{subfigure}{0.49\textwidth}
        \includegraphics[width=\textwidth]{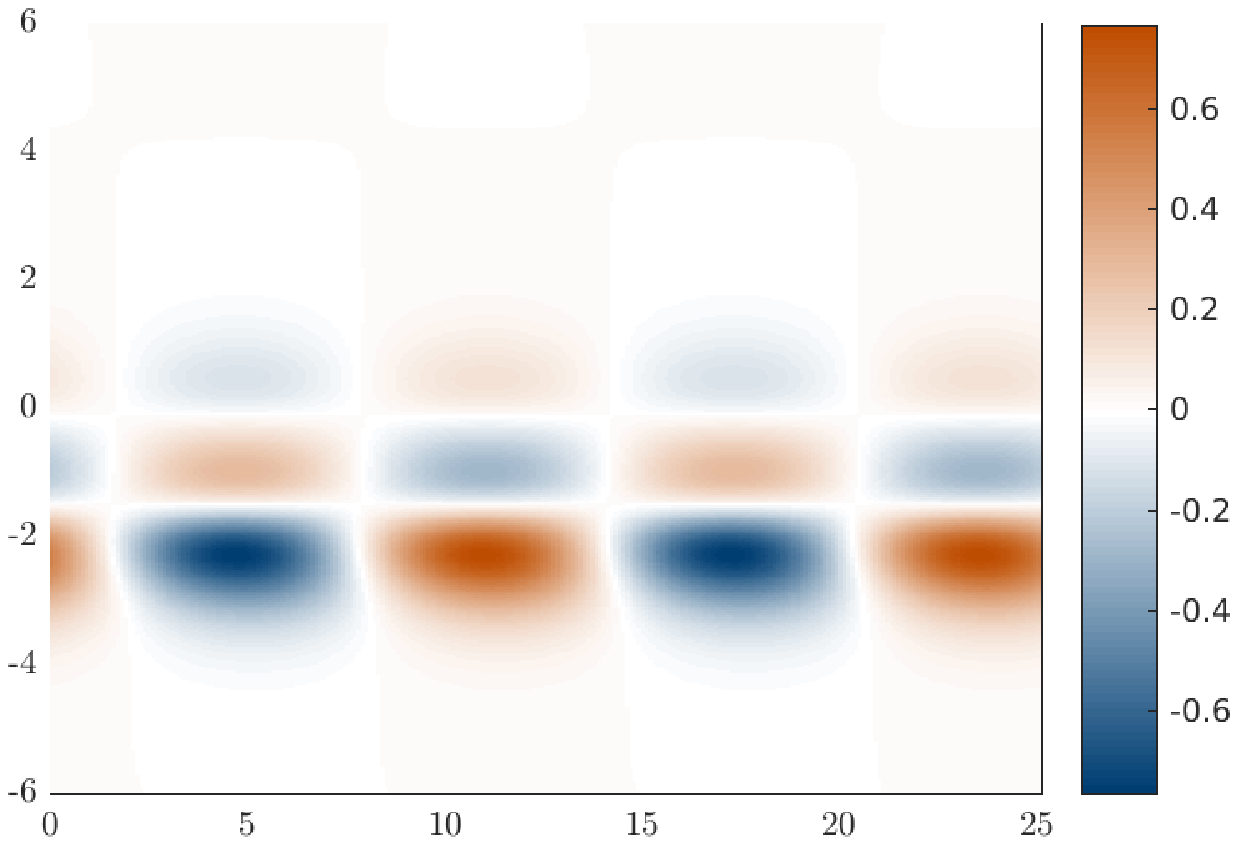}
        \raisebox{3.5cm}[0pt][0pt]{
            \parbox{3cm}{\caption{}\label{fig:instabreal}}
        }
        \end{subfigure}
        \begin{subfigure}{0.49\textwidth}
        \includegraphics[width=\textwidth]{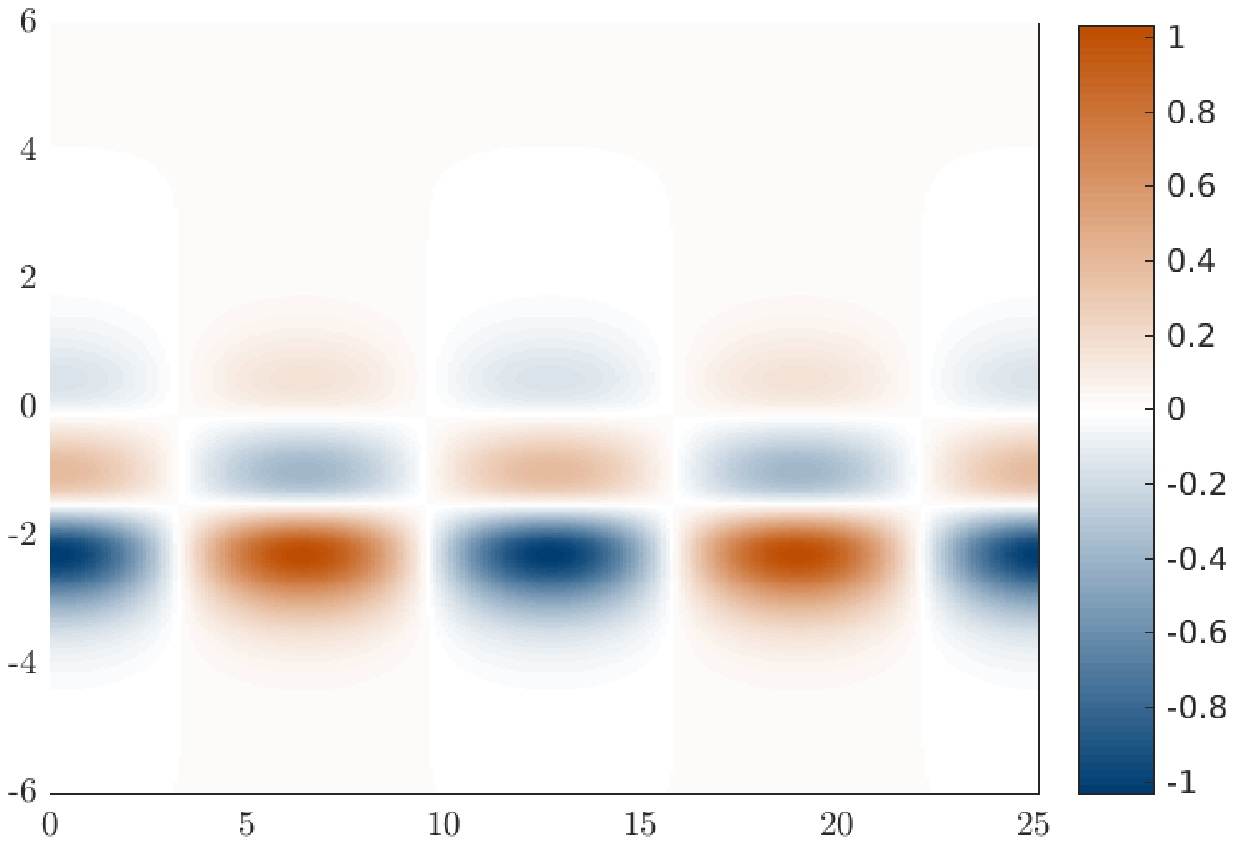}
        \raisebox{3.5cm}[0pt][0pt]{
            \parbox{3cm}{\caption{}\label{fig:instabimag}}
        }
        \end{subfigure}
        \vspace{-0.5cm}
        \caption{Real part of spanwise vorticity $\omega = \partial_x w - \partial_z u$ of the most unstable mode at $Ri_b=0.25$,
        for a flow with (a) $Re=4000$ and (b) $Re=40000$. Two domain lengths are shown horizontally. The full domain is $[-10,10]$ in the vertical direction.
        The growth rate of the $£Re=4000$ mode is $3.548\times10^{-6}+0.5229 i$. A critical layer is clearly visible near $z=-2$. An equivalent mode at critical layer
        $z=2$ also exists, with growth rate $3.548\times10^{-6}-0.5229 i$.
       }
                \label{fig:instability}
\end{figure}

A very weak linear instability, apparently hitherto unreported, is present in all states on
the bifurcation diagram. Figure \ref{fig:tanhstab} shows the maximum growth rate of linear instability
of the background state, as $Ri_b$ and $Re$ vary. For $Ri_b>1/4$, the new instability is the dominant one.
This has a phase speed of less than one, and manifests as convective rolls, advected through the domain,
above and below the interface at some critical layer, as shown in figure \ref{fig:instability}. As $Re\to\infty$, the growth rate tends to zero,
as required by the Miles-Howard theorem. Close agreement of growth rates, to one part in $10^3$, was found for this instability
 between the Arnoldi stability
algorithm of our code, and a direct solution of the Orr-Sommerfeld equations, using a MATLAB code
by W. D. Smyth.
Despite the small growth rate, at long times this instability leads to significant nonlinear behaviour
in the forced problem,
which eventually dominates and obscures all signature of KHI below $Ri_b=1/4$.
This means that the dynamical behaviour of KHI is very difficult to describe
in this model.

\subsection{Uniform stratification: the Drazin model}
\label{sec:linear}
We now consider the case with a uniform background stratification, so that $U=\tanh{z}$ but $B=z$.
This is also a commonly studied problem \citep{drazin_1958,churilov1987nonlinear,kaminski_transient_2014} as again,
linear stability analysis can be performed
analytically. As before, $Ri_m=Ri_b$ for this flow.
Linear stability analysis on a domain of infinite height
 tells us we should now take $L_x=2\sqrt{2}\pi$ to achieve $Ri_c\to 1/4$ as
$Re\to\infty$. As before, $L_z=10$. We use the same definition of energy $E$
as in the hyperbolic tangent case, though we note that only for uniform stratification does this form of
energy correspond precisely to the sum of perturbation kinetic and potential energies.

\begin{figure}
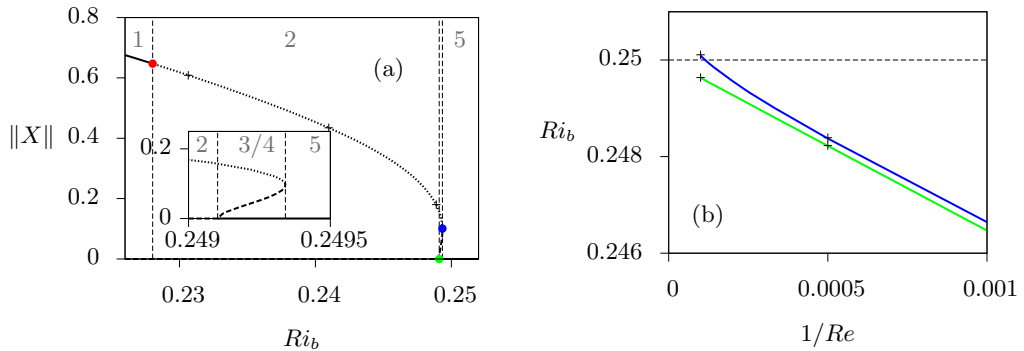

    \centering
    \begin{subfigure}{0.49\textwidth}
        \begin{gnuplot}[terminal=epslatex, terminaloptions={color colortext size 7cm,5cm}]
            set key off
            set size 1, 1
            set xtics out nomirror 0,0.01,0.25
            set ytics out nomirror 0,0.2,1
            set border lw 0.5
            set xrange [0.226:0.252]
            set yrange [0:0.8]
            set xlabel "$Ri_b$"
            set ylabel "$\\|X\\|$" offset 2,0 rotate by 0

            set style line 1 lw 3 lc 7

            set multiplot

            set label "1" at graph 0.02,0.9 tc rgb "#707070"
            set label "2" at graph 0.45,0.9 tc rgb "#707070"
            set label "5" at graph 0.93,0.9 tc rgb "#707070"

            plot \
            "-" w l lt 2 lw 1 lc rgb "black",\
            "-" w l lt 2 lw 1 lc rgb "black",\
            "-" w l lt 2 lw 1 lc rgb "black",\
             "fig3a/upperleft" w l ls 1 lt 1 lc rgb "black",\
             "fig3a/upperright" w l ls 1 lt 4 lc rgb "black",\
             "fig3a/midbranch" w l ls 1 lt 2 lc rgb "black",\
             "fig3a/lowerbranch" w l ls 1 lt 2 lc rgb "black",\
             "fig3a/basestate" w l ls 1 lt 1 lc rgb "black",\
             "fig3a/periodic" using 1:(sqrt(2*$2)) w lp ls 1 lt 2 lc rgb "red",\
             "fig3a/periodic" using 1:(sqrt(2*$3)) w lp ls 1 lt 2 lc rgb "red",\
             "-" pt 7 lc rgb "green",\
             "-" pt 7 lc rgb "blue",\
             "-" pt 7 lc rgb "red",\
             "-" w p pt 1 lc rgb "black"
             0.2280284762 0
             0.2280284762 0.8
             e
             0.2491026521 0
             0.2491026521 0.8
             e
             0.249341622 0
             0.249341622 0.8
             e
             0.2491026521 0
            e
            0.249341622	0.1008259683
            e
            0.2280284762 0.6472700882
            e
            0.24097788333892822266 0.43470163641812914479
            0.23065339028835296631 0.60868795682606313679
            0.24888950586318969727 0.18027666931680261242
            e

            set origin 0.12,0.2
            set size 0.6,0.5
            set xrange [0.249:0.2495]
            set yrange [0:0.25]
            unset xlabel
            unset ylabel
            unset label
            set label "2" at graph 0.05,0.82 tc rgb "#707070"
            set label "3/4" at graph 0.35,0.82 tc rgb "#707070"
            set label "5" at graph 0.85,0.82 tc rgb "#707070"

            set xtics out nomirror 0,0.0005,0.25 offset 0,0.5,0
            set ytics out nomirror 0,0.2,1 offset 0.5,0,0
            plot \
            "-" w l lt 2 lw 1 lc rgb "black",\
            "-" w l lt 2 lw 1 lc rgb "black",\
             "fig3a/upperleft" w l ls 1 lt 1 lc rgb "black",\
             "fig3a/upperright" w l ls 1 lt 4 lc rgb "black",\
             "fig3a/midbranch" w l ls 1 lt 2 lc rgb "black",\
             "fig3a/lowerbranch" w l ls 1 lt 2 lc rgb "black",\
             "fig3a/basestate" w l ls 1 lt 1 lc rgb "black"
             0.2491026521 0
             0.2491026521 0.8
             e
             0.249341622 0
             0.249341622 0.8
             e

        \end{gnuplot}
        \raisebox{4cm}[0pt][0pt]{
            \parbox{10cm}{\caption{}\label{fig:re1000lin}}
        }
        \vspace{-1cm}
    \end{subfigure}
    \begin{subfigure}{0.49\textwidth}
        \begin{gnuplot}[terminal=epslatex, terminaloptions={size 7cm,5cm}]
            set key off
            set xtics out nomirror 0,0.0005,0.001
            set ytics out nomirror 0,0.002,0.3
            set border lw 0.5
            set xrange [0:0.001]
            set yrange [0.246:0.251]
            set xlabel "$1/Re$"
            set ylabel "$Ri_b$" offset 3,0 rotate by 0

            f(x) = 0.249986-3.49354 * x**0.998637
            g(x) = 0.251093-0.319247 * x**0.625757

            plot \
            "-" w l lt 2 lw 1,\
            "fig3b.values" using (1/$1):3 w l lc rgb "blue" lw 3 lt 1,\
            "fig3b.values" using (1/$1):2 w l lc rgb "green" lw 3 lt 1,\
            "-" w p pt 1
           0 0.25
           0.002 0.25
           e
           0.0001 0.249638445477
           0.0001 0.25010872025156538623
           0.0005 0.248228756430
           0.0005 0.24839126111734705082
           e

        \end{gnuplot}
        \raisebox{2cm}[0pt][0pt]{
            \parbox{5cm}{\caption{}\label{fig:lintracking}}
        }
        \vspace{-0.5cm}
    \end{subfigure}
    \caption{(a) Bifurcation diagram of the flow with uniform background stratification, at $Re=4000$.
             The dashed vertical lines separate the numbered
             regions, as discussed in the text. Regions 3 and 4 are too small to label here.
             The solution branch has
              one unstable direction (\dashedline),
              two unstable directions (\dottedline),
              or is stable (\solidline).
              $Ri_h$ is plotted with a red dot, $Ri_s$ with a blue dot and $Ri_c$ with a green dot.
             (b) Variation of $Ri_c$ (green) and $Ri_s$ (blue) with $1/Re$. $Ri_s$ passes through $1/4$ at $Re\approx9000$.
             In both figures, the crosses mark points converged at the higher resolution of $256\times1024$.
             }
\end{figure}

Qualitatively, the bifurcation diagram is very similar to the $\tanh$ stratification case.
Figure \ref{fig:re1000lin} shows the diagram for $Re=4000$. The main difference from figure \ref{fig:re1000tanh}
is the lack of the first saddle-node bifurcation near the pitchfork. The values of the various bifurcation Richardson numbers are different,
for example the Hopf bifurcation at $Ri_h$ (shown in red) occurs at somewhat lower $Ri_b$ than before.
Also crucially, the
propagating linear instability described in section \ref{sec:tanh} is no longer present,
and consequently we can study the long-time behaviour of KHI.

The period of the Hopf bifurcation at $Re=4000$ is about $1690$ advective time units,
which is much too high to allow us
to converge the resulting periodic orbit directly, but long time integrations at a range of $Ri_b$ give us
an idea of the behaviour, since it appears to be stable in this case. Even this simple method becomes useless as
we approach $Ri_c$, since the period increases towards infinity. This is
the generic behaviour near a homoclinic bifurcation \citep{strogatz2018nonlinear}, which we believe
occurs somewhere between $Ri_c$ and $Ri_s$: the periodic orbit collides with the lower branch state.

\begin{figure}
    \begin{subfigure}{0.19\textwidth}
        \includegraphics[width=\textwidth]{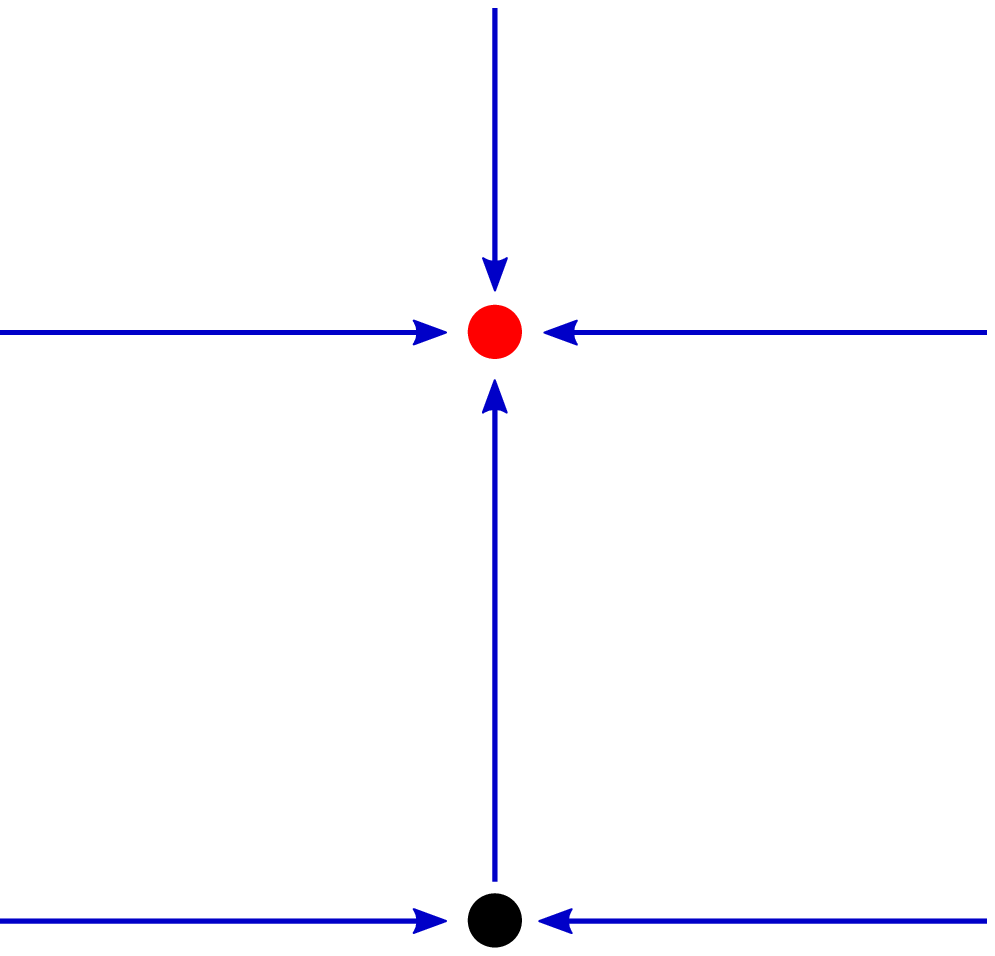}
        \caption{Region 1}
    \end{subfigure}
    \begin{subfigure}{0.19\textwidth}
        \includegraphics[width=\textwidth]{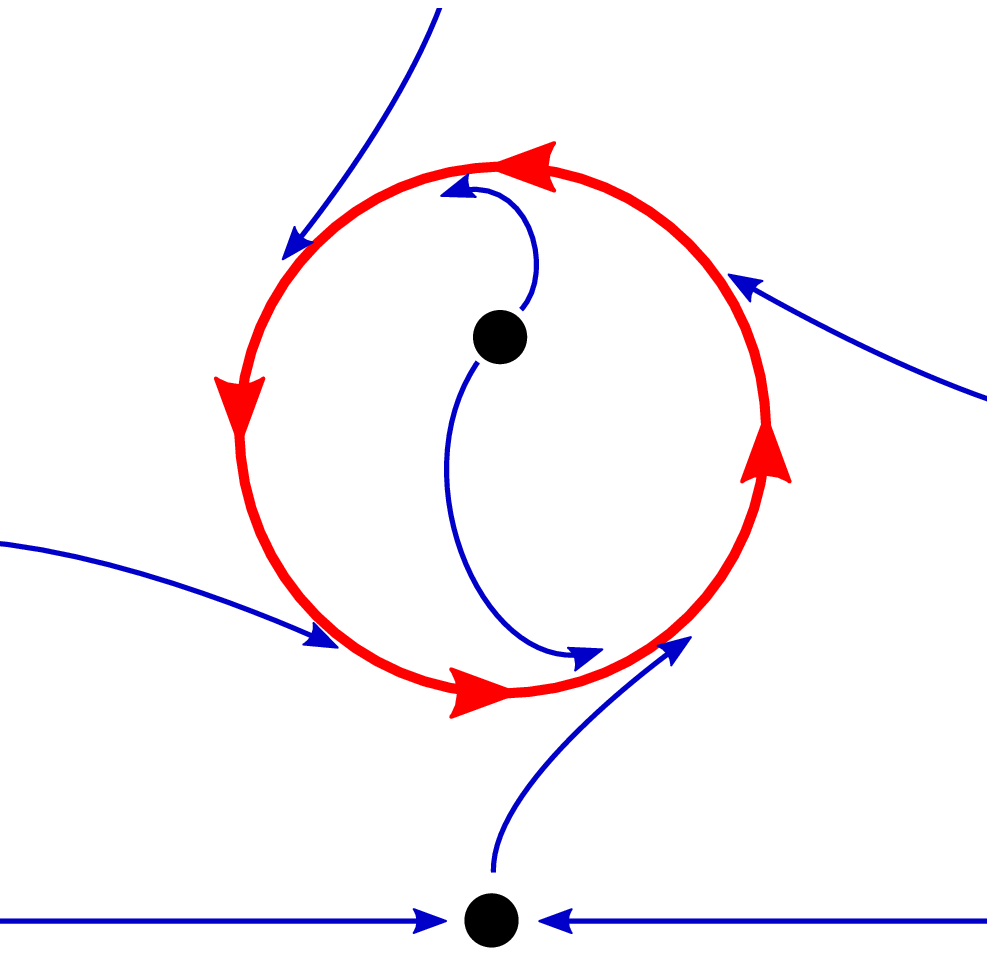}
        \caption{Region 2}
    \end{subfigure}
    \begin{subfigure}{0.19\textwidth}
        \includegraphics[width=\textwidth]{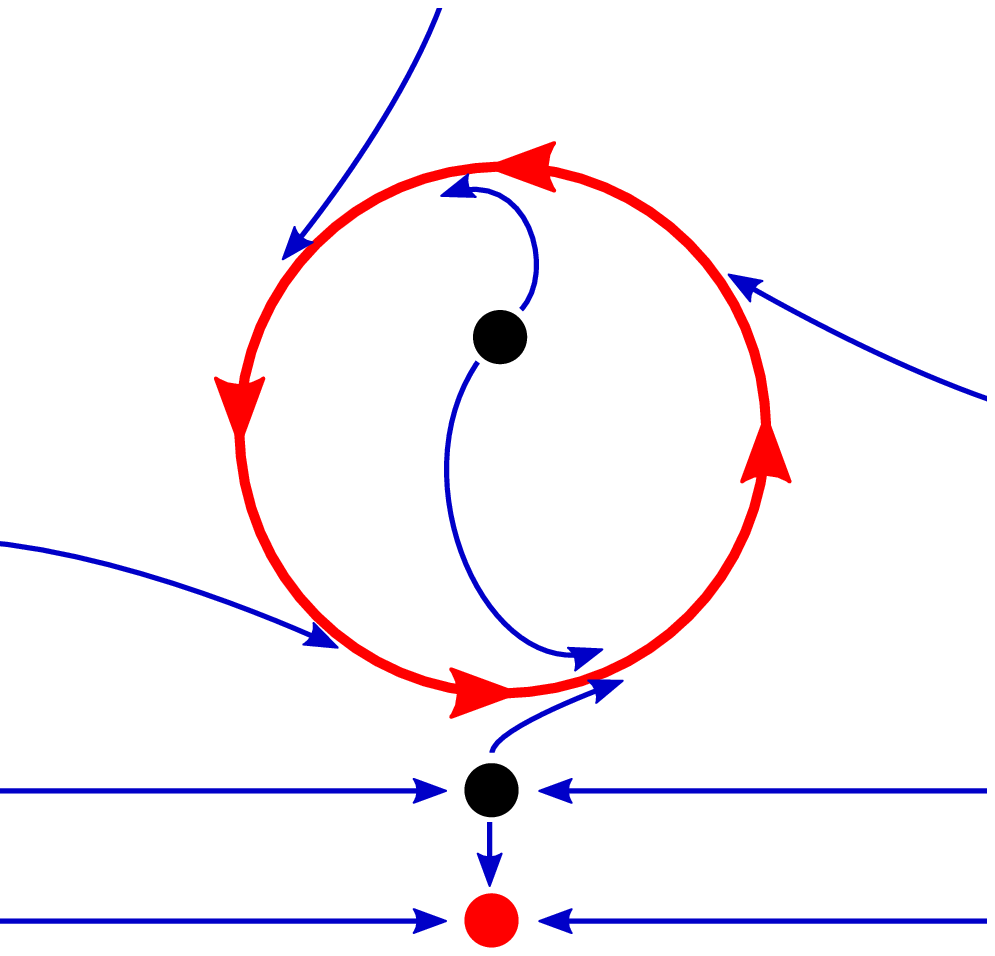}
        \caption{Region 3}
    \end{subfigure}
    \begin{subfigure}{0.19\textwidth}
        \includegraphics[width=\textwidth]{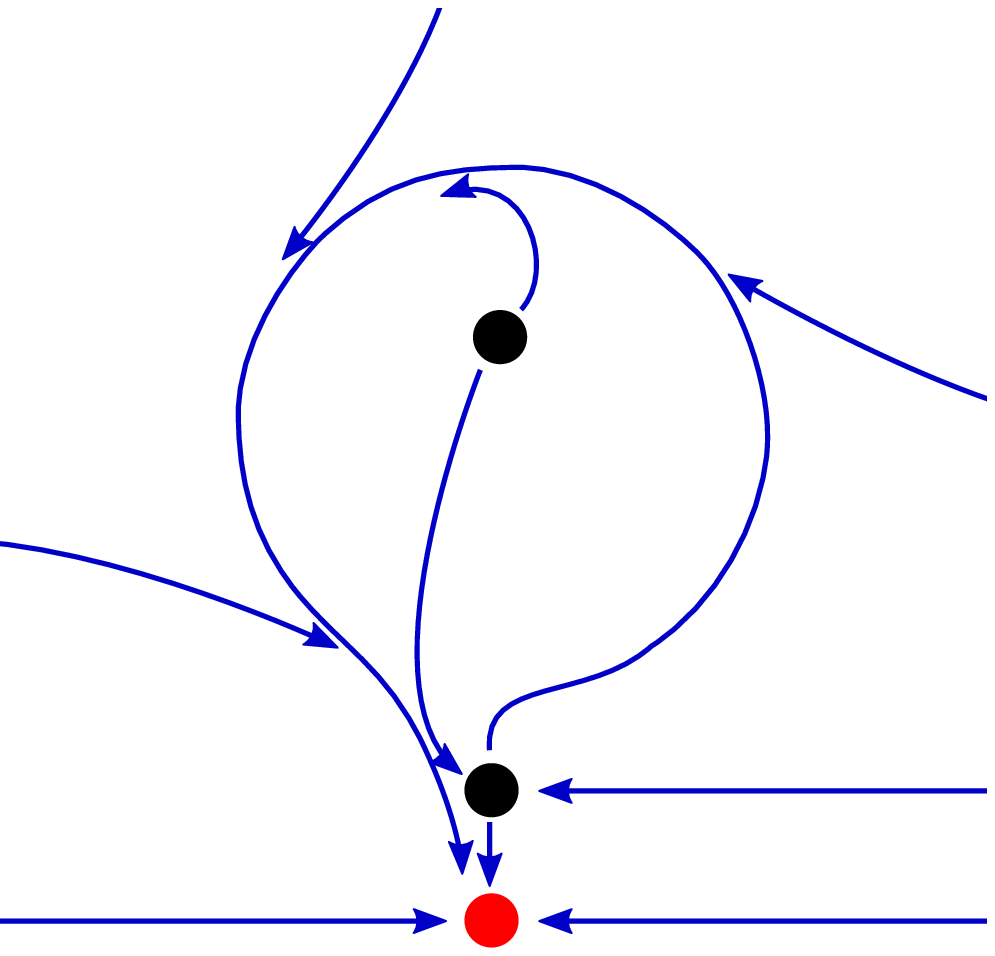}
        \caption{Region 4}
    \end{subfigure}
    \begin{subfigure}{0.19\textwidth}
        \includegraphics[width=\textwidth]{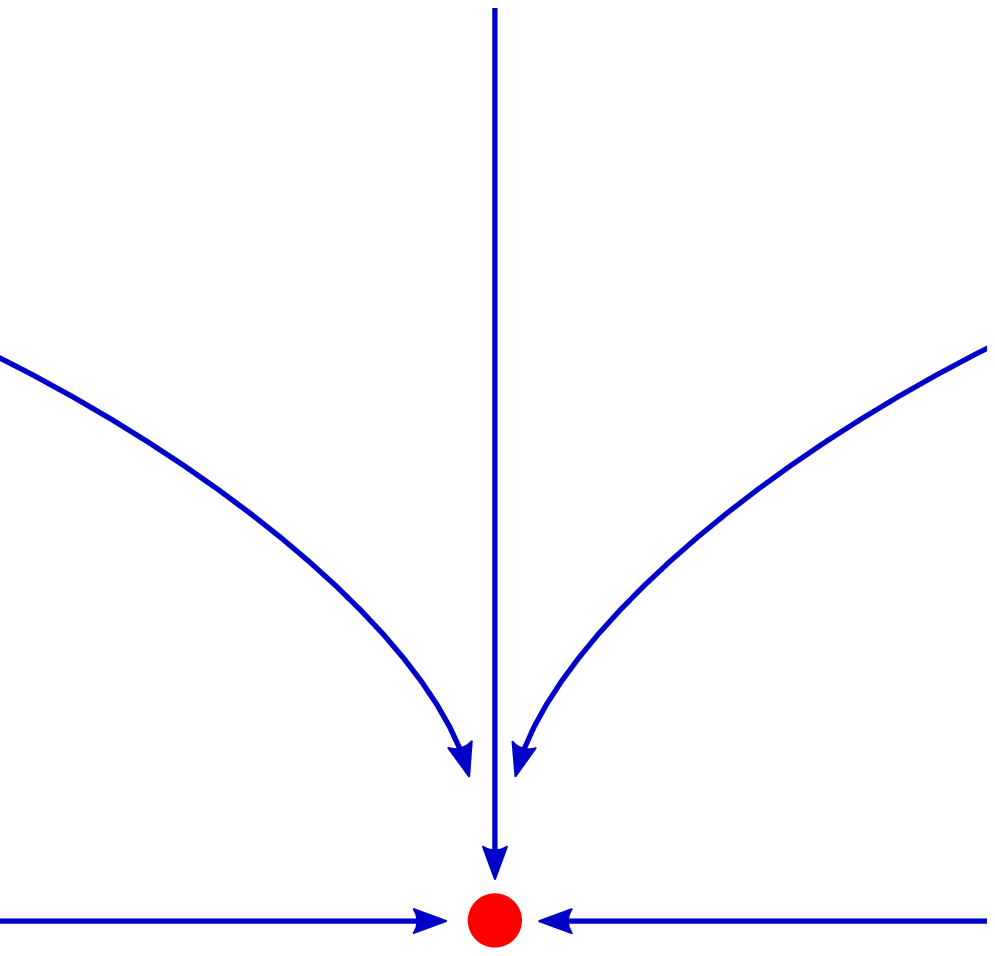}
        \caption{Region 5}
    \end{subfigure}
    \caption{Schematics of the dynamical system restricted to the two dimensional manifold of the two most
             unstable eigenmodes. The dots mark steady states, the lower being the laminar
             solution, and the lines show a few
             relevant trajectories. Solutions shown in red are stable, and those in black are unstable.}
    \label{fig:cartoons}
\end{figure}

The behaviour of the system, which is generic for sufficiently high $Re$, can be completely understood on a two-dimensional manifold described by the two most
unstable eigenmodes, as shown schematically in figure \ref{fig:cartoons}.
In region 1, where $Ri_b<Ri_h$, the laminar state
is unstable, and the instability saturates and eventually leads to the upper branch state, which is stable.
For $Ri_h<Ri_b<Ri_c$, region 2,
the laminar state and upper branch are both unstable, and perturbations lead to a stable periodic orbit.
Immediately above the pitchfork bifurcation $Ri_c$ in the region 3, the laminar state is stable and there exists a lower branch
edge state, which is unstable. If finite amplitude perturbations to the laminar state are past this edge,
they are attracted to the periodic orbit, and we have subcritical `transition'.
Region 4 is between the homoclinic bifurcation of the periodic orbit with the lower branch state,
and the saddle-node bifurcation of the lower and upper branches. Here, there are unstable finite amplitude states
and large transient trajectories, but the laminar state is the only attractor. In region 5,
past the saddle-node bifurcation, $Ri_b>Ri_s$, the laminar state is the only known exact coherent structure.
Of course, in reality the finite amplitude states break the translational symmetry of the laminar state,
and there are in fact a continuum of upper branch states, periodic orbits and so on, with a shift of origin.
Which of these the system is attracted to depends on the phase of the initial perturbation.

\begin{figure}
        \begin{subfigure}{0.49\textwidth}
        \includegraphics[width=\textwidth]{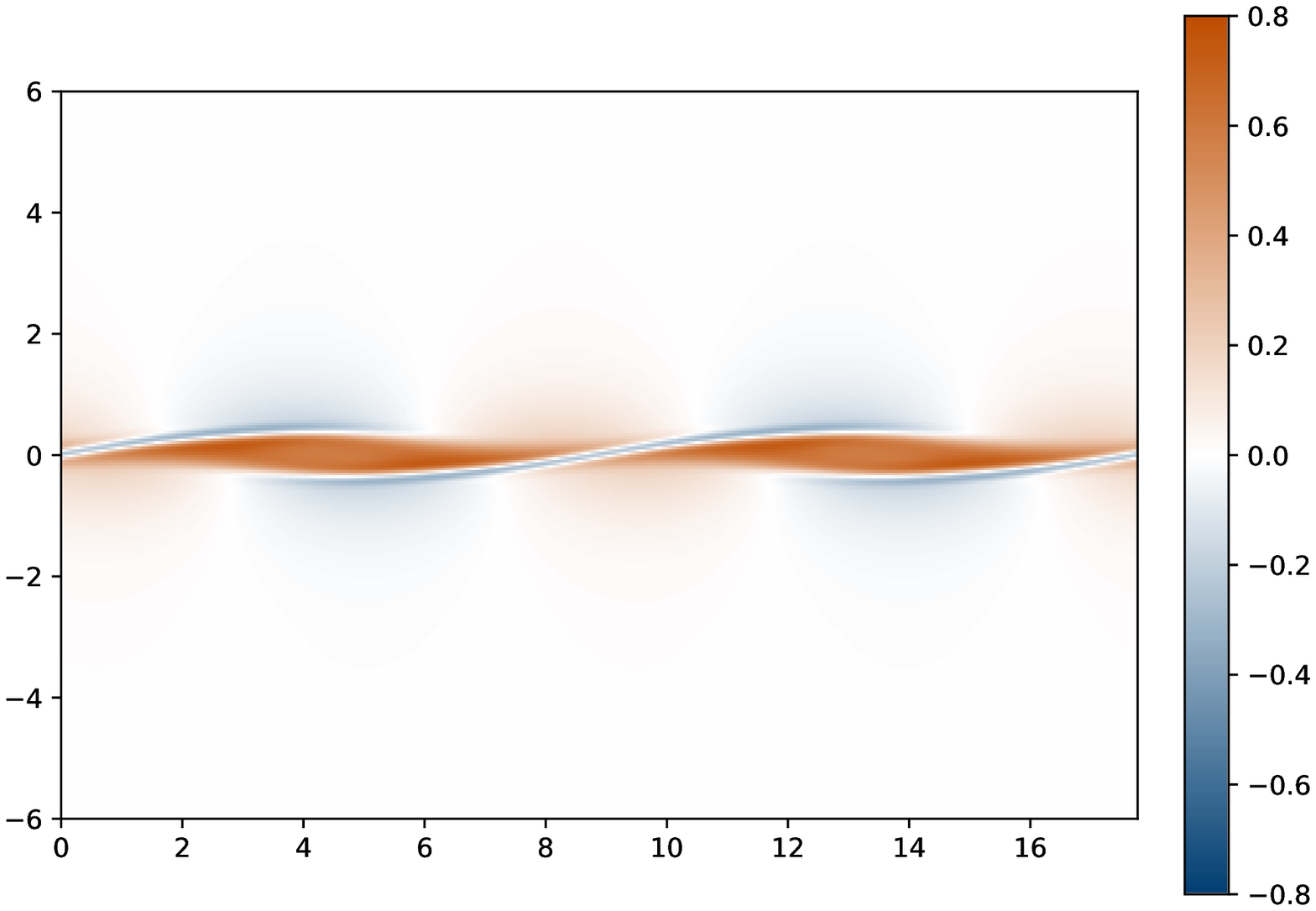}
        \raisebox{3.5cm}[0pt][0pt]{
            \parbox{3cm}{\caption{}\label{fig:vorthopf}}
        }
        \end{subfigure}
        \begin{subfigure}{0.49\textwidth}
        \includegraphics[width=\textwidth]{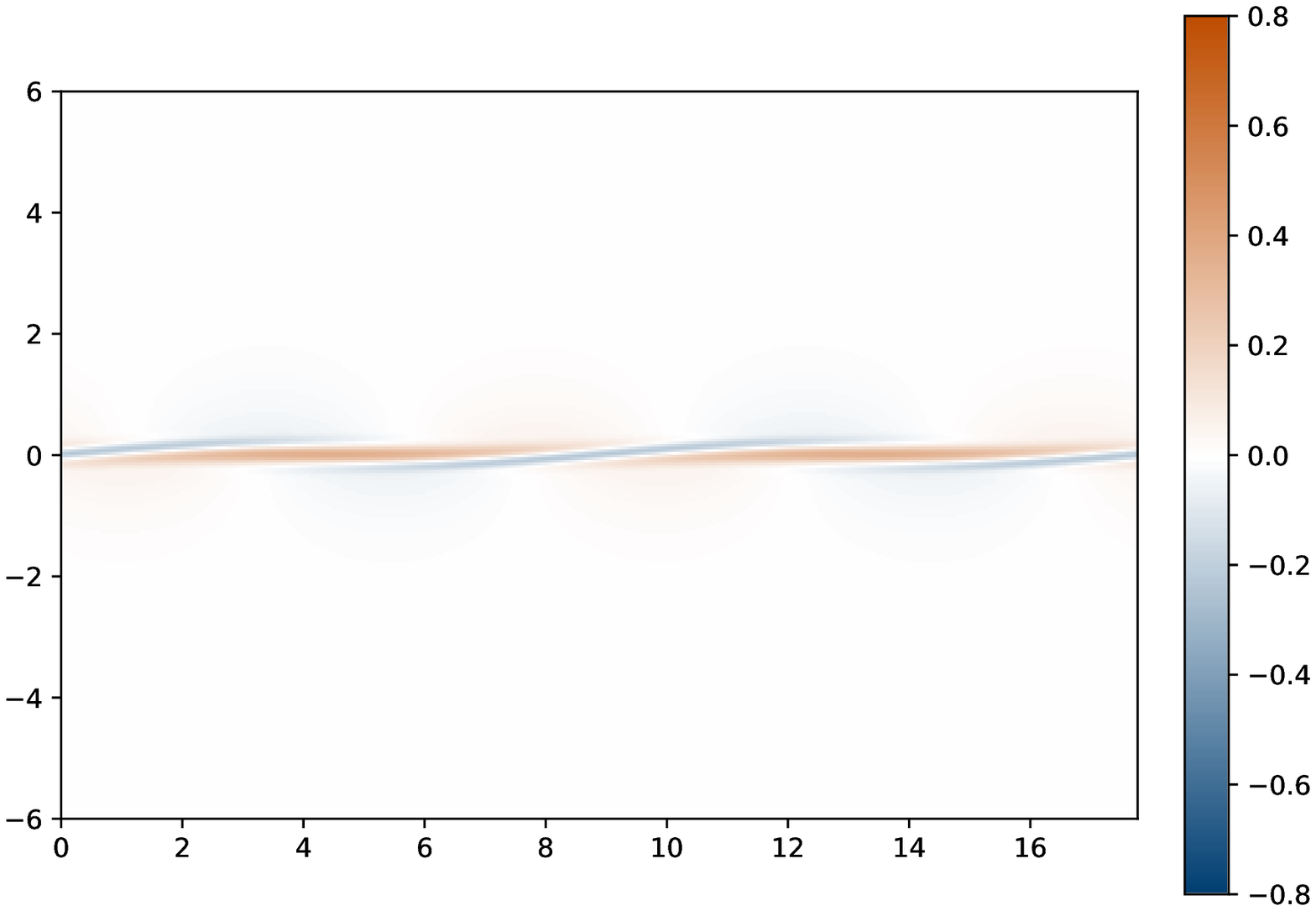}
        \raisebox{3.5cm}[0pt][0pt]{
            \parbox{3cm}{\caption{}\label{fig:vortsad}}
        }
        \end{subfigure}
        \vspace{-0.5cm}
    \caption{Spanwise vorticity $\omega = \partial_x w - \partial_z u$ of the stationary states at the (a) Hopf $Ri_h=0.22803$ and
             (b) saddle-node $Ri_s=0.24934$, for a flow with $Re=4000$. Two domain lengths are shown horizontally. The full domain is $[-10,10]$ in the vertical direction.}
             \label{fig:billows}
\end{figure}

Figure \ref{fig:billows} shows the vorticity structure of the steady states at two different values of $Ri_b$.
In the case of the Hopf bifurcation, billow-like structures are clearly
seen, bearing a strong resemblance to the saturated, unsteady billows found by \citet{howland_testing_2018}.
Increasing $Ri_b$ along the upper branch to the saddle-node bifurcation, these structures remain but become
significantly less pronounced. Baroclinic effects mean that the height of the billows decreases
with increasing $Ri_b$.

We track the values of $Ri_c$ and $Ri_s$ for $Re$ from $1000$ to $10000$ using the method described in
section \ref{sec:tracking}, and the results are shown on figure \ref{fig:lintracking}.
As $Re\to\infty$, extrapolation, assuming linearity in $1/Re$, suggests $Ri_c\to 0.25 -1.4\times10^{-5}$,
slightly less than $1/4$ because of the finite height
of the domain. Nevertheless, for $Re\gtrsim9000$, we find that $Ri_s>1/4$. It has been difficult to extrapolate $Ri_s$ convincingly,
though at $Re=\infty$, it seems likely that  $Ri_s\approx0.251$.
Since we have been
unable to find the location of the conjectured homoclinic bifurcation, we are unable to say whether
region 3, with a stable periodic orbit, extends above $Ri_b=1/4$, and hence whether the system is bistable
here. Nevertheless, region 4 certainly exists above $Ri_b=1/4$, so there will
be nonlinear transient behaviour, with the development of Kelvin-Helmholtz style billows as shown in
figure \ref{fig:billows}. We have also tracked the Hopf bifurcation
(omitted from figure \ref{fig:lintracking} for scale reasons)
and this shows a similar trend to the saddle-node bifurcation.



\section{Discussion and Conclusions}
\label{sec:conclusion}
The Miles-Howard theorem is an important result in the theory of linear stability of inviscid flows.
However, the fact it seems to work in more general conditions than those for which it is proven means
it has been informally applied as a `rule of thumb' at high $Re$.
We have shown that subcritical instability can exist in such flows, so that complex nonlinear behaviour
can occur even when the flow is linearly stable.
This is not a new result; \citet{maslowe1977} found subcritical instability in the Holmboe model
with $Pr=0.72$ and $Re=100$ using a weakly nonlinear analysis.
We note however, that this technique of finding the first order correction to the linear theory would have given misleading results
applied to the parameters we study, since in the Holmboe model, we find a saddle-node bifurcation very close to the pitchfork,
leading to subcritical instability instead of the apparent supercriticality.
Furthermore, the technique presented in this paper allows us to precisely find the location of the saddle-node bifurcations,
and demonstrate explicitly that finite ampltiude states exist at $Ri_b>1/4$, which has only been inferred previously.

We have been able to give a simple description of the dynamics in the Drazin model.
It is not immediately clear that the dynamics of the forced system studied here will be relevant to
those of an unforced system, which has traditionally been used as a model for geophysical flows.
The incredibly long periods of the orbits born from the Hopf bifurcation
discussed earlier, for example, mean that in an unforced problem, the background flow would have diffused
almost entirely away before one complete cycle.
 Nevertheless, the instability of the unforced flows still leads to saturated
states very similar to the steady solutions we have found, and the subcriticality we have demonstrated
would certainly lead to nontrivial transient behaviour.

Despite our results, it seems that $Ri_b=1/4$ is indeed a useful `rule of thumb' for stability in physical flows.
The subcriticality we have found extends only very slightly about $1/4$ in both cases studied.
However, this is in apparent disagreement with the results of \citet{maslowe1977}, who found large subcriticality.
This is likely because of the different value of $Pr$ studied. Indeed, in a follow-up work \citet{brown1981}
found subcriticality when $Pr<1$ but supercriticality when $Pr>1$, and showed that higher order terms must be considered at our choice of $Pr=1$,
so we would like to extend this work to the more oceanographically relevant range $Pr\sim O(10)$.

In addition to these finite amplitude nonlinear states, we have found linear instability with $Ri_b>1/4$ in the Holmboe
model (see figure \ref{fig:instability}), which disappears
as $Re\to\infty$, as required by the Miles-Howard theorem. A similar phenomenon was found by \citet{miller_lindzen_1988}.
 However, their instability had large growth rates and required a carefully constructed flow.
We have found an instability in a widely used model, hitherto unreported to the best of our knowledge.
The new instability has a tiny growth
rate at physically realistic $Re$. This suggests
it can be ignored in oceanic problems, but does not entirely explain why it has not been discussed before.
It is commonly assumed that finite $Re$ effects are always stabilising compared to inviscid behaviour.
This instability demonstrates that such assumptions should be checked carefully.
While it is not appropriate to classify this instability as `classic' Holmboe wave instability,
since it lacks the characteristic `wave-interaction' resonance between an interfacial gravity wave and vorticity waves
localised at the edge of the shear layer, we conjecture that it may be homotopically connected to Holmboe instability as parameters are varied,
since it has a similar phase speed and occurs at similar values of $Ri_b$. This is an area for future research.

\bibliographystyle{../jfmlatex/jfm}
\bibliography{pck19}

\end{document}